\documentclass[]{raa}            
\usepackage{graphicx,times}
\usepackage{natbib}
\usepackage{graphicx}
\usepackage{epstopdf}
\usepackage{color}

\providecommand{\bm}[1]{\mbox{\boldmath$#1$\unboldmath}}

\begin{document}

   \title{The LAMOST Survey of Background Quasars in the Vicinity Fields of the M\,31 and M\,33 -- III. Results from the 2013 Regular Survey}

 \volnopage{ {\bf 2015} Vol.\ {\bf X} No. {\bf XX}, 000--000}
   \setcounter{page}{1}

   \author{Zhi-Ying Huo
      \inst{1}
   \and Xiao-Wei Liu
      \inst{2}
    \and Mao-Sheng Xiang
      \inst{2}
   \and Jian-Rong Shi
      \inst{1}       
   \and Hai-Bo Yuan
      \inst{3,5}
   \and Yang Huang
      \inst{2}
   \and Yong Zhang
      \inst{4}
  \and Yong-Hui Hou
      \inst{4} 
  \and Yue-Fei Wang
      \inst{4}    
  \and Ming Yang
      \inst{1}     
   }


   \institute{National Astronomical Observatories, Chinese Academy of Sciences,
             Beijing 100012, P.R. China; {\it Email: zhiyinghuo@bao.ac.cn}  
        \and  Department of Astronomy, Peking University,
              Beijing 100871, P.R. China
       \and
             Kavli Institute for Astronomy and Astrophysics, Peking University,
             Beijing 100871, P.R. China    
       \and  Nanjing Institute of Astronomical Optics \& Technology, 
             Chinese Academy of Sciences, Nanjing 210042, P.R. China
       \and  LAMOST Follow      
\\
\vs \no
   {\small Received [2015] [March] [31]; accepted [2015] [May] [11] }
}

\abstract{In this work, we report new quasars discovered in the vicinity fields of the Andromeda (M\,31)
and Triangulum (M\,33) galaxies with the LAMOST (Large Sky Area Multi-Object Fiber
Spectroscopic Telescope, also named Guoshoujing Telescope) during the 
2013 observational season, the second year of Regular Survey.   
In total, 1330 new quasars are discovered in an
area of $\sim$133 deg$^2$ around M\,31 and M\,33.  
With $i$ magnitudes ranging from 14.79 to 20.0, redshifts from 0.08 to 4.85, the 1330 new quasars 
represent a significant increase of the number of identified quasars in the vicinity fields 
of M\,31 and M\,33.
Up to the moment, there are in total 1870 quasars discovered by LAMOST in this area (see also Huo et al. 2010, 2013). 
The much enlarged sample of known quasars in this area can
potentially be utilized to construct a precise astrometric reference frame for the
measurement of the minute proper motions of M\,31, M\,33 and the associated substructures,
vital for the understanding of the formation and
evolution of M\,31, M\,33 and  the Local Group of galaxies.
Meanwhile, amongst the sample, there are in total 45, 98 and 225 quasars with $i$ magnitudes 
brighter than 17.0, 17.5 and 18.0 respectively. In the aforementioned brightness bins, 
15, 35 and 84 quasars are reported here for the first time, 
6, 21 and 81 are reported in Huo et al. (2010, 2013), while 0, 1 and 6 are from the Sloan 
Digital Sky Survey, and 24, 41 and 54 are from the NED database. 
These bright quasars provide an invaluable sample for the kinematics and chemistry study of the
interstellar/intergalactic medium of the Local Group. 
}

   \authorrunning{Z.~Y. Huo et al.}            
   \titlerunning{Background Quasars in the Vicinity Fields of M\,31 and M\,33} 
   \maketitle


%
%
\section{Introduction}           
\label{sect:intro}

Since the first optical spectral identifications of a quasar (Schmidt 1963), the number of known quasars
has increased steadily and rapidly. In particular, the Sloan Digital Sky Survey (SDSS;
York et al. 2000) has hither to discovered about $\sim$260,000 quasars (see Schneider et al. 2010; 
Paris et al. 2012, 2014 and references therein). Being the most energetic objects
in the universe, quasars have been widely used as excellent tracers to study a variety of astrophysical
problems, such as large scale structure of the universe (e.g. Boyle et al. 2000), 
 central massive black holes of galaxies (e.g. Corbett et al. 2003), galaxies
formation and evolution (e.g. Gebhardt et al. 2000), and the interstellar/intergalactic (ISM/IGM) medium 
(e.g. Murdoch et al. 1986, Savage et al. 2000). 

The Andromeda galaxy (M\,31) is the most luminous member of the Local Group of galaxies and 
the nearest archetypical spiral galaxy that serves as one of the best astrophysical 
laboratories for the studies of the formation and evolution of galaxies.  
The deep photometric surveys of Canada-France-Hawaii Telescope 
(CFHT, Ibata et al. 2007; McConnachie et al. 2009) 
have revealed a whole variety of complex substructures within hundreds kilo-parsec (kpc) of M\,31,
some extending all the way to the Triangulum galaxy (M\,33),  pointing toward
 a possible encounter of the two galaxies in the past.
Further chemical and kinematic investigations of M\,31, M\,33 and associated substructures
are vital for the assemblage history of the Local Group. 

With a distance of $\sim$ 785 kpc (McConnachie et al. 2005), M\,31 is moving towards 
the Milky Way at a speed of 117 km s$^{-1}$ (Binney \& Tremaine, 1987, p.605).
Meanwhile, its transverse velocity remains unmeasurable for a long time. 
The distance and proper motion (PM) of M\,33, have been measured by the Very Long 
Baseline Array (Brunthaler et al. 2005).
Nevertheless, water maser sources in M\,31 have been discovered only recently 
(Darling 2011),  so do not have long enough time baseline for an accurate PM measurements yet. 
Based on the kinematics of M\,33 and IC\,10, two satellite galaxies of M\,31, 
Loeb et al. (2005) and van der Marcel \& Guhathakurta (2008) present theoretical estimates 
of the PM of M\,31, about 80 km s$^{-1}$ (20 $\mu$as yr$^{-1}$).
Based on {\it Hubble Space Telescope (HST)} imaging data spanning 5--7 years, 
Sohn, Anderson \& van der Marel (2012) present the first direct PM measurements 
of three M\,31 fields, each of $\sim$ 2.7$\times$2.7 arcmin$^2$.  
Given the sparse of known background quasars, compact background galaxies are used instead as
reference sources for the PM measurements. 
Given the importance of PM measurements for understanding the kinematics of M\,31, M\,33 and those 
associated structures,  securing a sufficiently large number and density of background quasars 
would be extremely useful to construct an accurate reference frame for precise PM measurements.  
 
A series studies based on {\it The Spectroscopic and Photometric Landscape of AndromedaÕs 
Stellar Halo} (SPLASH; see Dorman et al. 2012; Gilbert et al. 2009, 2012, 2014; 
Tollerud et al. 2012 and reference therein) survey present the kinematical and 
chemical properties of stars in the M\,31 inner and extended halos, as well as in features of tidal debris, 
and dwarf galaxies. Spectroscopic measurements of the chemical composition of stars at the M\,31 distance
are not easy tasks even for ten-meter class telescopes, including KECK.
Absorption-line spectroscopy of bright quasars in the vicinity fields of M\,31 and M\,33
can be used to probe the distribution, kinematics and chemical
composition of the ISM/IGM associated with the Milky Way, M\,31, 
M\,33, and related substructures. 
Several studies published hitherto (see Savage et al.\ 2000; Schneider et al.\ 1993
and references therein), based on data collected with the {\it Hubble Space Telescope 
Quasar Absorption Line Key Project}, have revealed a wide range of  ionization states, 
chemical compositions and kinematics of the Milky Way halo gas. 
Rao et al. (2013) study the properties of the extended gas
halo of M\,31 using {\it HST} Cosmic Origin Spectrograph (COS) observations of 10 background quasars.
Such studies are however currently limited to a few lines of sight, restricted by the available
number of bright (low-redshift) quasars in the fields of interest.   

The Large Sky Area Multi-Object Fiber Spectroscopic Telescope (LAMOST) is an innovative reflecting 
Schmidt telescope with a design allowing both a large light-collecting aperture (the effective aperture 
varying between 3.6 m -- 4.9 m, depending on the declination and hour angle of pointing) 
and a wide field of view of 5 degree in diameter
(Cui et al. 2012; Wang et al. 1996; Su \& Cui 2004). With a spectral resolution of R $\approx$ 1800 and a wavelength 
coverage of 3700 -- 9000 ${\rm \AA}$, LAMOST can simultaneously record spectra of up 
to 4000 celestial objects.  
After two years' commissioning (2009-2010), one year's Pilot Survey  (2011),
the LAMOST Regular Survey was initiated in October 2012 (Zhao et al. 2012).

This is the third installment presenting the results of LAMOST survey of background quasars in 
the vicinity fields of M\,31 and M\,33.
In the first of this series, we reported 14 new quasars discovered with the LAMOST
using the early commissioning data collected in 2009 (Huo et al. 2010, hereafter Paper I).
In the second paper, we presented 526 newly discovered quasars, 
based on data collected during the 2010 commissioning 
as well as the 2011 Pilot Survey (Huo et al. 2013, hereafter Paper II).
In this paper, we present newly discovered background quasars in this area
in the 2013 observational season, the second year of the Regular Survey.


\section{Candidate Selection} 
\label{sect:QSOcand}

Quasar candidates are selected based on the optical and near infrared (IR) 
photometric data available in the vicinity fields of M\,31 and M\,33.
Within the SDSS coverage, low-redshift quasar candidates are selected by following 
Richards et al. (2002). Quasar candidates and stars are well separated in the SDSS
color-color diagrams. Candidates down to an $i$-band magnitude limit of 20.0\,mag 
(with a surface number density of $\sim$27 candidates per square degree) are selected 
(see Papers I and II for more details). 
In the central area of M\,31 and M\,33, no reliable photometric data are provided by the SDSS
as well as the Xuyi Schmidt Telescope Photometric Survey (XSTPS; Liu et al. 2014),
quasar candidates are selected using the data from the Kitt Peak National Observatory (KPNO) 4m 
telescope survey of M\,31 and M\,33 (Massey et al. 2006), after transforming the
KPNO $UBVRI$ magnitudes to SDSS $ugriz$ magnitude using the transformation of Jester et al. (2005)
deduced for quasars of redshifts $z \leq 2.1$. 

\begin{figure}[h!!!]
\centering
\includegraphics[width=12.0cm,angle=0]{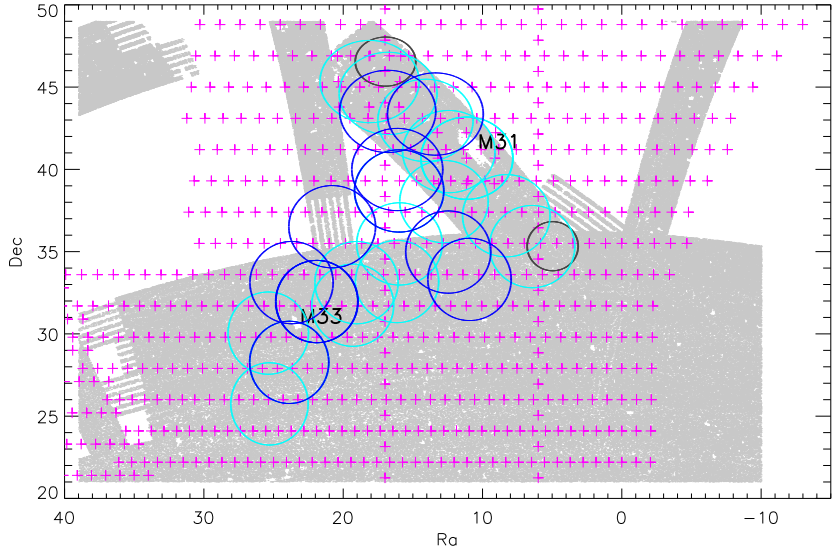}
\caption{Sky coverage of the SDSS and XSTPS in the M\,31 and M\,33 area.
The positions of M\,31 and M\,33 are marked. The grey area represents the SDSS sky coverage of this area. 
The magenta plus symbols delineate the individual field centers of the XSTPS  M\,31/M\,33 survey, 
with each field covering 2$\times$ 2 deg$^2$. 
The black circles represent the three spectroscopic fields targeting quasar candidates in the outer 
halo of M\,31 by SDSS,  
the one in the northeast area represents two fields of the same central positions.
The cyan circles labeled the LAMOST fields observed in 2010 and 2011, respectively. 
The blue circles represent the LAMOST fields observed in the 2013 observational season. 
}
\label{qsocand}
\end{figure}

In addition to the optically selected candidates, the Wide-field 
Infrared Survey Explorer (WISE; Wright et al. 2010) sources of colors $W1-W2 \geq 0.8$  
are selected as candidates (Yan et al. 2013, Stern et al. 2012). 
The IR sources are cross-correlated with the available optical point-source catalogs of SDSS, KPNO and XSTPS (see Paper II for more details). By this methods, 
about 26 candidates per square degree are selected down to a limiting magnitude of $i \le 20.0$\,mag.
As shown in Wu et al. (2012), this IR-based
selection is capable of finding quasars of redshifts up to $z < 3.5$.
Fig.\,\ref{qsocand} shows the sky coverage of SDSS and XSTPS in the M\,31 and M\,33 area.
 
\section{Observations and Data Reduction}
\label{sect:Obs}

The quasar candidates in the vicinity fields of M\,31 and M\,33 have been targeted by the LAMOST 
since its initial commissioning in 2009, see Paper I and Paper II for the previous results of 
quasars discovered with LAMOST in this area.  Here we report results based on data 
collected in the 2013 observational season, the second year of the LAMOST Regular 
Survey. Through two years (2009-2010) of commissioning, one year (2011) of Pilot Survey and
one year (2012) of Regular Survey we have seen the performance of LAMOST been stabilized, 
with some steady improvement over time (Huo et al. 2015, in prep.; Yuan et al. 2015, in prep.).

There are 10 plates with quasar candidates observed by LAMOST in the 2013 observational season. 
Similar to the LAMOST Spectroscopic Survey of the Galactic Anti-center (LSS-GAC; Liu et al. 2014, 
Yuan et al. 2015 for detail), 
three categories of spectroscopic plates are designated, bright (B), medium (M) and faint (F), 
targeting sources of different brightness for efficient usage of observing time of different qualities 
and at the same, avoiding fiber cross-talking.
All quasars candidates are assigned a fiber in the M plates only regardless of their brightness, 
except for a few quasar candidates of $r$ magnitude brighter than 16.3. The latter have been assigned
a fiber in B plates and observed repeatedly.  The exposure times for the individual M plates 
vary between 1800 to 2400s, repeated 2 or 3 times for cosmic ray rejections and for 
buildup of the signal-to-noise ratio (SNR). 
The spectra have a wavelength coverage of  3700-9000 ${\rm \AA}$, with a resolving power
of R $\approx$ 1800.
In total, 7425 unique quasar candidates were targeted in the 2013 season, some of them 
were targeted repeatedly. Because of the circular field of view of LAMOST,  
some field overlapping is necessary in order to have a contiguous sky coverage.
See Fig.\,\ref{qsocand} for the fields targeted by LAMOST in 2013.  

The spectra were reduced with the LAMOST two-dimensional (2D) pipeline of version 2.6 
(Luo et al. 2004, 2012; Luo et al. 2015).
Given the relatively low throughputs of LAMOST at  
blue wavelengths, and the relatively faintness of quasar candidates, flux-calibration often
induces larger uncertainties. As such, for the purpose of identifying quasars of interest here, 
we use spectra without flux calibration.  

\section{Results and Discussion}
\label{sect:results}

The current work is based on data collected during the 2013 second year 
Regular Survey of LAMOST.
The performance of LAMOST, including throughput and fiber positioning, improved
significantly comparing to the earlier time of operation of LAMOST. Sky subtraction,
although improved, is still not as good as expected, especially at wavelengths 
longer than 7200 ${\rm \AA}$.
Fortunately, quasars are easily identified for their characteristic broad emission lines.
We require that at least one emission line is securely identified.
We visually examined the one-dimensional (1D) extracted spectra of quasar candidates one by one, 
and this led to the identifications of 1545 quasars in the 10 observed fields near M\,31 and M\,33, 
and 1370 of them are newly discovered. If we adopt a luminosity cut 
by following the SDSS Quasar Survey (Schneider et al. 2010, see also the references therein), 
with an absolute $i$-band magnitude of $M_i=-22.0$ in a cosmology $H_{\rm 0} = 70$ km s$^{\rm -1}$
Mpc$^{\rm -1}$, $\Omega_{\rm M} = 0.3$, and $\Omega_{\rm \Lambda}  = 0.7$,
the number of identified quasars is 1503, of which 1330 are new.

\begin{table}[h!!!]
\bc
\begin{minipage}[]{150mm}
\caption[]{Catalog of New Quasars in the Vicinity Fields of M\,31 and M\,33 Discovered by LAMOST 
in the 2013 observational season.}
\label{tab_cat}
\end{minipage}
\setlength{\tabcolsep}{2pt}
\small
 \begin{tabular}{cccccccccccccc}
  \hline\noalign{\smallskip}
Object  & R.A. & Dec. (J2000) & Redshift &$u$ & $g$ & $r$ & $i$ & $z$ & A(i)  &Selection$^*$   \\  
  \hline\noalign{\smallskip}

  J003234.36+335838.0   &  8.143199  &  33.977227   &  1.46  &  20.16  &  20.10  &   19.82  &  19.64  &  19.62   &   0.17  &    W,S,-  \\
  J003237.68+330028.2   &  8.157018  &  33.007835   &  1.22  &    -       &  19.64  &  19.54   &  19.16  &      -        &   0.19  &    W,-,-   \\
  J003238.14+331814.1   &  8.158927  &  33.303941   &  1.57  &  19.82  &  19.41  &   19.35  &   19.09 &   19.08  &   0.16  &    W,S,-  \\
  J003326.08+322235.3   &  8.358672  &  32.376485   &  2.87  &  21.08  &  20.00  &   19.80  &   19.65 &   19.49  &   0.16  &    W,-,-   \\
  J003333.87+331441.5   &  8.391137  &  33.244884   &  2.51  &  19.83  &  18.98  &   18.50  &   18.40 &   18.19  &   0.17  &    W,-,-   \\
  J003347.28+340456.8   &  8.447031  &  34.082462   &  2.22  &    -        &  19.48  &   19.25  &   19.24  &     -      &   0.20  &    W,-,-   \\
  J003354.78+335628.9   &  8.478290  &  33.941377   &  1.72  &  21.34  &  21.05  &   20.46  &   19.95  &  19.89  &   0.17  &   W,-,-   \\
  J003437.21+334634.1   &  8.655059  &  33.776146   &  1.72  &  18.16  &  18.24  &   18.00  &   17.62  &  17.76  &   0.16  &   W,S,-  \\
  J003445.88+333018.4   &  8.691207  &  33.505132   &  2.22  &     -       &  18.98  &   19.19  &   18.79  &     -       &   0.17  &   W,-,-   \\
  J003459.55+323110.9   &  8.748138  &  32.519718   &  0.44  &  19.60  &  19.22  &   19.05  &   18.94  &  18.82   &  0.16  &    W,S,-   \\
    
   \noalign{\smallskip}\hline
\end{tabular}
\ec
\tablecomments{1.0\textwidth}{$^*${\it W} represents those candidates selected by the IR criteria 
based on the WISE data, {\it S} represents optically selected using the SDSS criteria 
for low-redshift quasars, {\it M} represents optically selected from the KPNO 
data (Massey et al. 2006). 
Only a portion of Table is shown here for illustration. The whole Table 
contains information of 1330 new quasars is available in the online electronic version.}
\end{table}

\begin{figure}[htbp]
\centering
\includegraphics[width=14.5cm,angle=0]{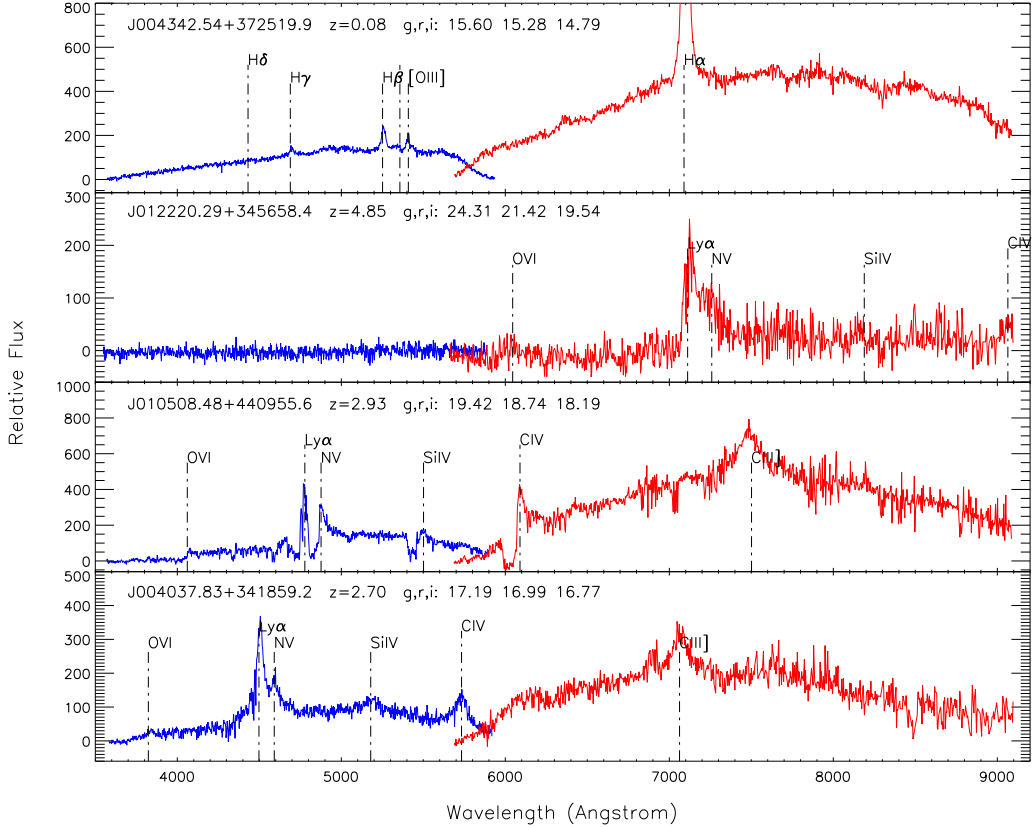}
\caption{Example of newly discovered quasars by LAMOST in the 2013 Regular survey.
The vertical dash-doted lines labeled identified emission lines.
labeled. The $y$-axis is the relative flux in units of counts per pixel. The spectra have been binned by
a factor of 4, and cosmic rays remained after pipeline processing removed manually for clarity.
}
\label{spectra}
\end{figure}

All the new quasars have reasonable SNRs and at least one emission line securely identified, 
allowing reliable redshift estimations (see Fig.\,\ref{spectra}).  
Table\,\ref{tab_cat} present the catalog of these newly discovered quasars, including 
target designation (in format of J{\it hhmmss.ss+ddmmss.s}), J2000.0 coordinates 
(right ascension and declination in decimal degrees), redshifts, 
the observed SDSS $u$, $g$, $r$, $i$ and $z$ magnitudes without extinction corrections, 
$i$-band extinction from Schlegel et al. (1998), also the selection criteria used. 
Only a portion of the Table is shown here. The whole Table
containing information of 1330 new quasars is available in the online electronic
version only. 
In the last column of the Table\,\ref{tab_cat}, {\it W} indicates targets
selected with the IR criteria based on the WISE data, {\it S} indicates targets
selected using the SDSS criteria for low-redshift quasars, and {\it M} indicates 
those targets selected based on the KPNO data of the Local Group of galaxies.
No quasars selected from the KPNO data are discovered in this 2013 observational season. 
The Columns 5-9 list the SDSS $u$, $g$, $r$, $i$ and $z$ magnitudes. 
For some targets,  the $u$ and $z$ magnitudes are not given, those targets are 
selected by cross-matching the WISE-IR candidates
with the XSTPS optical catalog.  

Fig.\,\ref{spectra} shows the LAMOST spectra of 4 new quasars with a wide 
range of properties. The spectra have not been flux-calibrated. 
At wavelengths shorter than 4000 ${\rm \AA}$ where the
instrument throughputs are relatively low, as well as between the dichroic cross-over
wavelength range 5700--6000 ${\rm \AA}$, the spectra have relatively low SNRs.  
Since the LAMOST spectra are oversampled, they are binned by a factor of 4 to
improve the SNRs.  Cosmic rays remained after pipeline processing
have been removed manually for clarity. 

Here we present a brief description of the properties of the 4 quasars shown in Fig.\,\ref{spectra}. 
At a redshift of 0.08 and with an $i$ magnitude of 14.79, 
J004342.54+372519.9 is the brightest and of the lowest redshift quasar in the catalog reported
in this paper.
With a redshift of 4.85, J012220.29+345658.4 is the highest redshift quasar here. 
It is also the highest redshift quasar discovered by LAMOST by the end of the 2013 
observational season, as far as we know. 
This object is also observed with YFOSC (Yunnan Faint Object Spectrograph and Camera) 
mounted on the Lijiang Gaomeigu 2.4m telescope of the Yunnan Astronomical Observatory
in January of 2015. The identification and redshift of the LAMOST spectrum are confirmed. 
J010508.48+440955.6 is an example of quasars with broad absorption lines in our catalog,
with an emission line redshift of 2.93.
J004037.83+341859.2, at a redshift of $z=2.70$ and an $i$-band absolute magnitude of $M_i = -30.19$\,mag, 
is the most luminous quasar in our catalog.

\subsection{Spacial Distribution}

\begin{figure}[hbtp]
\centering
\includegraphics[width=13.5cm,angle=0]{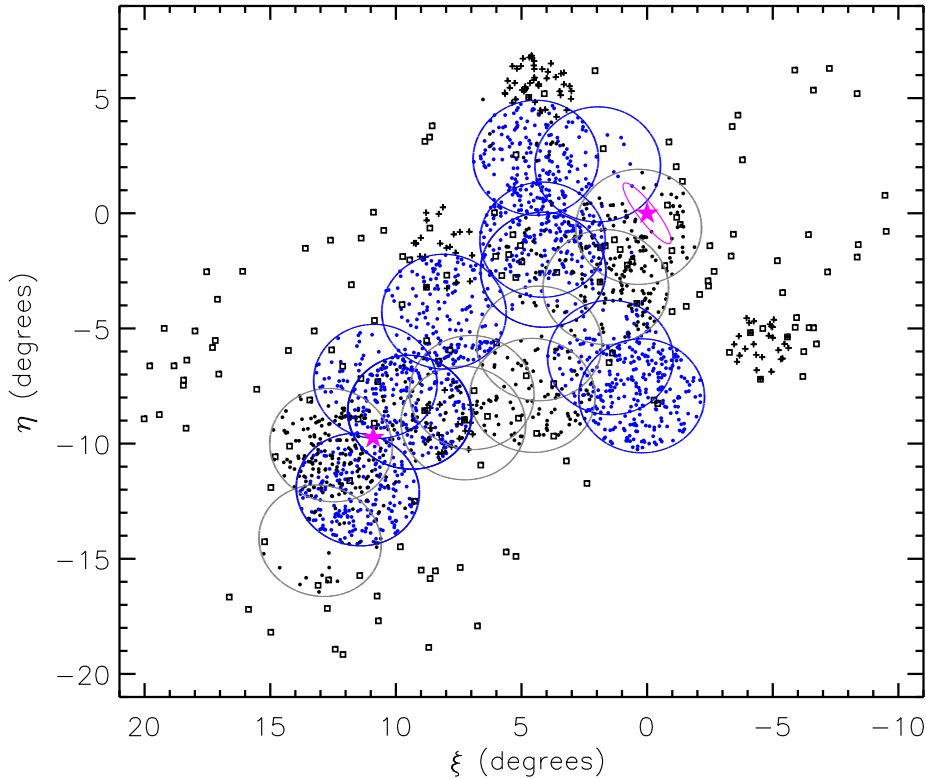}
\caption{
Spatial distributions of background quasars in the vicinity fields of M\,31 and M\,33. 
Blue filled circles represent quasars identified by LAMOST in the 2013 dataset.
Black filled circles represent quasars identified in the LAMOST 2009, 2010, and 2011 datasets. 
Pluses and open squares represent SDSS quasars and previously known quasars with 
redshifts given in the NED archive, respectively.
The central positions of M\,31 and M\,33 are marked by the magenta stars,  
while the magenta ellipse represents the optical disk of M\,31 of radius $R_{25} = 95\farcm3$. 
The big grey and blue circles represent LAMOST fields observed in 2011 and 2013, respectively.
}
\label{fig:spatial}
\end{figure}

The spatial distribution of all known background quasars in the vicinity fields of M\,31 and M\,33
are shown in Fig.\,\ref{fig:spatial}, in the $\xi$--$\eta$ plane.
Here $\xi$ and $\eta$ are respectively the Right Ascension and Declination 
offsets relative to the optical center of M\,31 (Huchra, Brodie \& Kent 1991).
There are, in total, 1870 quasars discovered by LAMOST by the end of the
2013 observational season, including 1330 newly discovered quasars in an area of 
$\sim$133 deg$^2$ reported in
this paper, 14 and 526 quasars discovered using the LAMOST earlier
observations and Pilot Survey, as reported in Papers I and II, respectively. 
In addition, 75 quasars are identified in three SDSS spectroscopic 
plates in two fields in the outer halo of M\,31 along its major axis 
(Adelman-McCarthy et al. 2006, 2007), 43 quasars are discovered
serendipitously in two {\it Sloan Extension for Galactic Understanding 
and Exploration} (SEGUE; Yanny et al. 2009) plates in this area, and 155 previously known quasars 
with redshifts listed in the NED archive within a 10 degree of M\,31 and of M\,33.
In Fig.\,\ref{fig:spatial}, the central positions of M\,31 and M\,33 are marked by magenta stars. 
The magenta ellipse represents the optical disk of M\,31, with an optical radius 
$R_{25} = 95\farcm3$ (de Vaucouleurs et al. 1991), an inclination angle $i = 77$\,$^\circ$ degree 
and a position angle $PA$ =35\,$^\circ$ (Walterbos \& Kennicutt 1987).  
The number of known quasars behind the extended halo and associated substructures of M\,31 
has increased by a substantial amount. The much enlarged number of known quasars 
in the vicinity fields of M\,31 and M\,33 will serve as an invaluable sample for future 
PMs and ISM/IGM studies of M\,31, M\,33 and the Local Group.  

The LAMOST quasars are mainly distributed around M\,31 and M\,33, as well as
the area between M\,31 and M\,33, including the Giant Stellar Stream (see Fig.\,\ref{fig:spatial}).
However, the spatial distribution of identified quasars by LAMOST is not uniform. 
Of the 10 plates observed in 2013, 5 plates have 200 or more
quasars identified, one plate with only 75 quasars identified, while the remaining
4 plates have a number of identified quasars between 100 and 200.  
The highest density of quasars identified by LAMOST in 2013 is $\sim$12.8 quasars per square degree.
The overall efficiency of quasar identifications in the 2013 observational season is $\sim$21\% (1545/7425),  although the improvement is significant comparing to what achieved in Papers I and II, 
this is still lower than that reported by Stern et al. (2012) for candidates selected using the 
WISE colors.
We believe the low yields are mainly caused by the poor observing conditions or the non-uniform 
performance of the 4000 fibers of LAMOST.

\subsection{Magnitude and Redshift Distributions}
    
The magnitude limit of this sample is $i=20.0$\,mag.  Amongst the 1330 newly discovered
quasars, there are 15/35/84 new quasars with $i$-band magnitude brighter than
17.0/17.5/18.0\,mag, respectively. 
Amongst the quasars discovered using the commissioning and Pilot Survey observations (Papers I and II),
there are 6/21/81 quasars brighter than the aforementioned magnitude limits. 
Amongst the 75
quasars discovered by SDSS in the three spectroscopic plates in the outer halo of M\,31 
(Adelman-McCarthy et al. 2006, 2007) and the 43 serendipitously discovered quasars 
in the two SEGUE plates of this field, the corresponding numbers are 0/1/6.
And there are 24/41/54 quasars brighter than 17.0/17.5/18.0\,mag $i$-magnitudes respectively,
within 10 degree of M\,31 and of M\,33 in the NED archive (see Paper II).
There are in total 45/98/225 quasars with $i$ magnitudes brighter than 17.0, 17.5 and 18.0\,mag
in this area.
These bright quasars will be an invaluable sample to probe the properties of ISM/IGM 
in M\,31/M\,33 and the Local Group.

\begin{figure}[htbp] 
   \centering
   \includegraphics[width=13.5cm, angle=0]{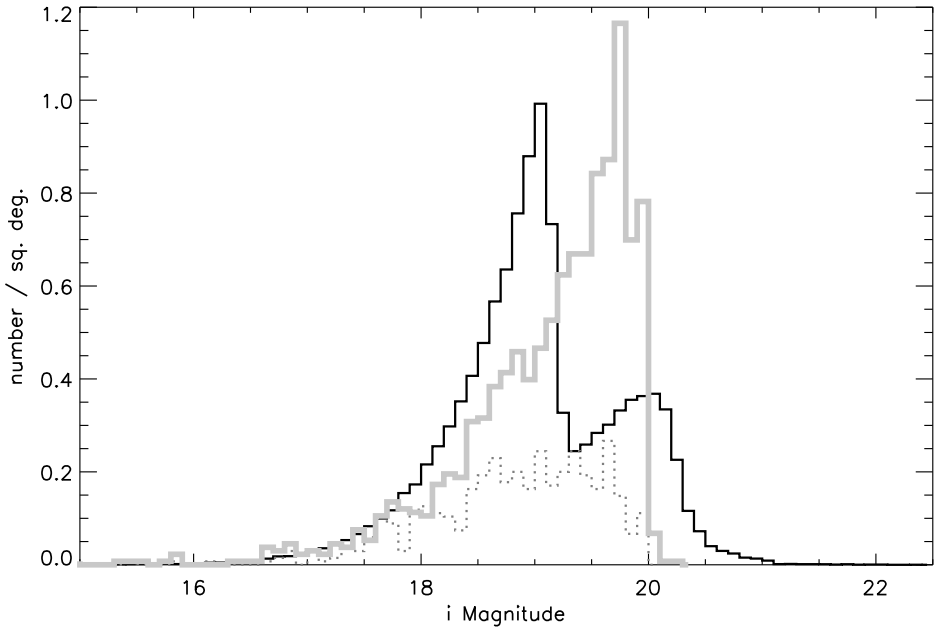} 
   \caption{Histogram distribution of $i$-band magnitudes of the 1503 quasars detected by 
   LAMOST in 2013 (grey thick line),
   including 1330 newly discovered and 173 previously detected by LAMOST. 
   For comparison, the distribution of the 532 quasars detected by LAMOST in 2011 (grey dotted line),
   and that of SDSS DR7 quasars (Schneider et al. 2010; black solid line) are also plotted for comparison.
   The magnitude bin size is set to 0.1\,mag. The y-axis gives the quasar number density per square degree.
   }
   \label{fig:mag}
\end{figure}

Fig.\,\ref{fig:mag} shows the magnitude distribution of the 1503 quasars detected 
in the 2013 datasets, including the 1330 newly discovered and 173 previously known 
but re-observed by LAMOST. The magnitude has a bin size of 0.1\,mag, the y-axis is given in
quasar number density. 
For comparison, we also plot the $i$-band magnitude distribution of the 532 quasars
detected by LAMOST in 2011 (Paper II), and the magnitude distribution of 
SDSS DR7 quasars  (Schneider et al. 2010). 
We can see that there is a significant improvement in quasar number density by LAMOST in
2013 comparing to that of 2011.  It is difficult to compare our distribution directly with that of SDSS,
given the different target selection algorithms. It is however clear
that there are still a number of quasars remained to be identified by LAMOST.
This is also suggested by the relatively 
low quasar identification efficiency by LAMOST, as pointed out above.  

\begin{figure}[h!!!] 
   \centering
   \includegraphics[width=13.5cm, angle=0]{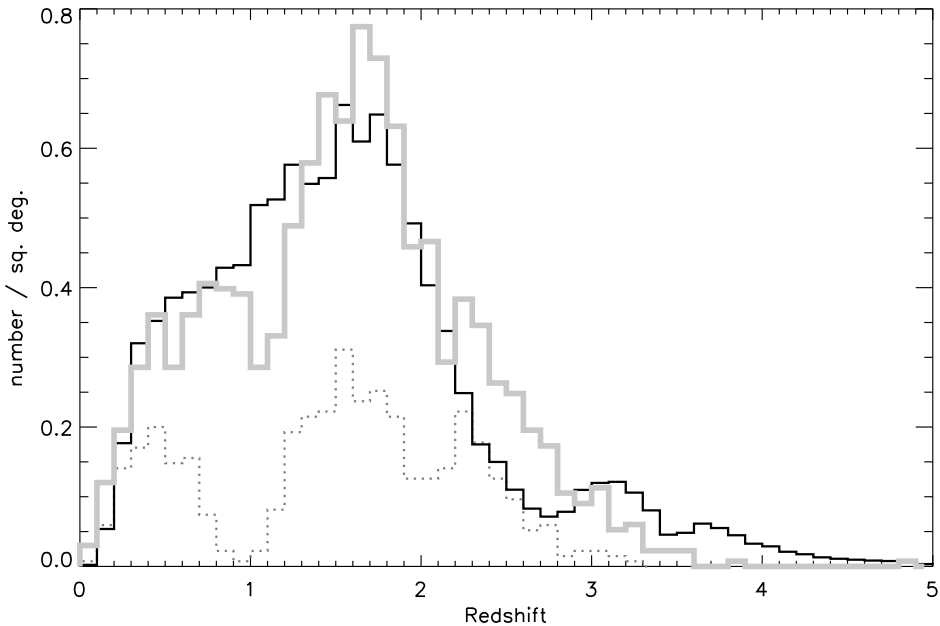} 
   \caption{Redshift distribution of the 1503 quasars discovered by LAMOST in 2013 (grey thick line).
   The grey dotted line represents the redshift distribution of the 532 quasars discovered by LAMOST in 2011. 
   The redshift distribution of the SDSS DR7 quasars (Schneider et al. 2010; black line) is also shown for comparison.
   The relative low number density at $z \sim 2.7$ seen in the distribution of SDSS quasars is caused by the 
   degeneracy of colors of stars and quasars around this redshift.
   The redshift bin size is set to 0.1. The y-axis gives the quasar number density per square degree.}
   \label{fig:redshift}
\end{figure}

\begin{figure}[h!!!] 
   \centering
   \includegraphics[width=13.5cm, angle=0]{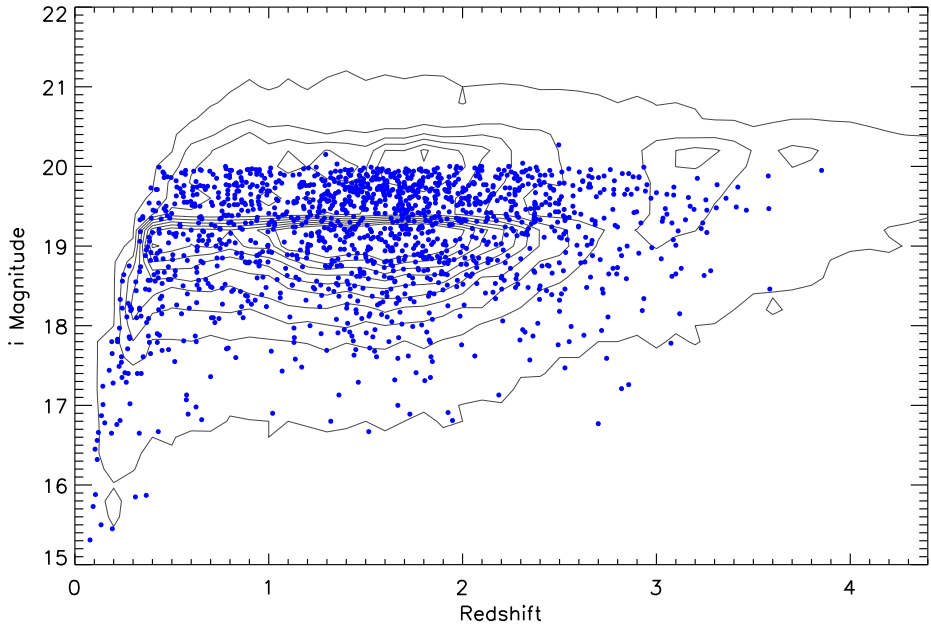} 
   \caption{Redshift versus $i$-band magnitude distribution of the 1503 LAMOST detected 
   quasars in 2013 (blue points). 
   The grey contours represent the redshift versus $i$-band magnitude distribution of the SDSS DR7 
   quasars (Schneider et al. 2010). 
   The steep gradients near $i \approx 19.1$ and $i \approx 20.2$\,mag are due to the magnitude limits of 
   the SDSS selection algorithms for low- and high-redshift quasars, respectively. 
   The relative low number density at $z \sim 2.7$ is 
   caused by the degeneracy of the SDSS colors of stars and quasars near this redshift.}
   \label{fig:mag-redshift}
\end{figure}

Fig.\,\ref{fig:redshift} shows the redshift distribution of the 1503 quasars
detected by LAMOST in 2013. 
The redshift has a bin size of 0.1. The y-axis gives quasar number density per square degree, 
as in Fig.\,\ref{fig:mag}. The spectra of quasars with the lowest and highest redshift, J004342.5+372519.9
and J012220.3+345658.4, are shown in Fig.\,\ref{spectra}.  
The second highest redshift is $z=3.85$, and there are 41 quasars with redshifts higher 
than 3.0 in this new catalog.
In Fig.\,\ref{fig:redshift}, we also plot the redshift distribution of the 532 quasars 
detected by LAMOST in 2011 (Paper II), and the redshift distribution of 
SDSS DR7 quasars (Schneider et al. 2010), for comparison.
We can see that there is a significant improvement in the number density of quasars detected by LAMOST in
2013 compared to that of 2011. In some redshift ranges, for example between 
2.0 and 3.0, the number density of quasars identified by LAMOST
is relatively higher than that of SDSS DR7. This is due to the different target selection algorithms adopted.
Our targets include candidates selected with the 
WISE IR colors, thus increasing the probability of finding quasars of redshifts up to $z < 3.5$ (Wu et al. 2012).
For redshifts of $z \leq 1.3$ or $z \geq 3.0$, the number density of quasars identified by 
LAMOST is lower than that of SDSS.  
The selection effects in the redshift distribution of quasars detected 
in 2013 is not as severe as that of 2011. 
The distribution of quasars discovered in 2011 has two obvious troughs in redshift,
one near redshift $\sim$ 0.9 and another near $\sim$ 2.1 (see Paper II for details). 
This improvement is benefitted from improvement in the LAMOST performance as well as in the 2D pipeline.
Fig.\,\ref{fig:mag-redshift} shows the distribution of quasars in the redshift-magnitude plane detected by 
LAMOST in 2013. The distribution concentrates near redshifts between  
1.2 and 2.0, and $i$ magnitudes between 19.0 and 20.0. This is also apparent in Fig.\,\ref{fig:mag} 
and Fig.\,\ref{fig:redshift}.

Given the heterogeneous target selection algorithms and the relatively low identification efficiency,  
the sample is not intended for the purpose of statistical analysis.

\normalem
\begin{acknowledgements}

This work is supported by National Key Basic Research Program of China 2014CB845705, 
and NSFC grant 11403038. 
We thank the staff of Li-Jiang Observatory, Yunnan Observatories of China for 
the observation and technological support, particularly Jin-Ming Bai and Wei-Min Yi,
also help from Hong-Liang Yan from National Astronomical Observatories, Chinese Academy of Sciences.
The Guoshoujing Telescope (the Large Sky Area Multi-Object Fiber Spectroscopic
Telescope; LAMOST) is a National Major Scientific Project built by the Chinese
Academy of Sciences. Funding for the project has been provided by the National
Development and Reform Commission. The LAMOST is operated and managed by the
National Astronomical Observatories, Chinese Academy of Sciences.  
 
Funding for the SDSS and SDSS-II has been provided by the Alfred P. Sloan
Foundation, the Participating Institutions, the National Science Foundation,
the U.S. Department of Energy, the National Aeronautics and Space
Administration, the Japanese Monbukagakusho, the Max Planck Society, and the
Higher Education Funding Council for England.  The SDSS Web Site is
http://www.sdss.org/.

This publication makes use of data products from the Wide-field Infrared Survey
Explorer, which is a joint project of the University of California, Los
Angeles, and the Jet Propulsion Laboratory/California Institute of Technology,
funded by the National Aeronautics and Space Administration.

This research has made use of the NASA/IPAC Extragalactic Database (NED) which
is operated by the Jet Propulsion Laboratory, California Institute of
Technology, under contract with the National Aeronautics and Space
Administration.

\end{acknowledgements}


\end{document}